\documentclass[pdflatex,sn-nature,oneside]{sn-jnl}

\usepackage{amsmath,amssymb}
\usepackage{graphicx}
\usepackage{bm}
\usepackage{booktabs}
\usepackage{makecell}
\makeatletter
\@twosidefalse
\@mparswitchfalse
\makeatother
\begin{document}

\title[MLCD prediction of defect formation energies]{Predicting large-supercell defect formation energies from machine-learning charge density models trained on small supercells}

\author*[1]{\fnm{Junjie} \sur{Zhou}}
\author*[1]{\fnm{Menglin} \sur{Huang}}\email{menglinhuang@fudan.edu.cn}
\author*[1]{\fnm{Shiyou} \sur{Chen}}\email{chensy@fudan.edu.cn}

\affil*[1]{\orgdiv{College of Integrated Circuits and Micro-Nano Electronics, and Key Laboratory of Computational Physical Sciences (MOE)}, \orgname{Fudan University}, \orgaddress{\city{Shanghai}, \postcode{200433}, \country{China}}}

\abstract{First-principles defect calculations are often limited by the cost of the large supercells required to suppress image interactions. Machine-learning interatomic potentials (MLIPs) provide another alternative, but training defect MLIPs typically requires thousands of structures and weeks of data generation. Since charge density is the key to density-functional-theory (DFT), we propose a machine-learning charge density (MLCD) route for predicting defect formation energies with higher data efficiency. We optimize the training set by integrating small supercells of varying sizes for better extrapolation, allocating their proportions based on spatial charge-density analysis. With only 96 supercells containing 16--96 atoms as the dataset, MLCD accurately predicts the formation energies of four intrinsic defects in 360-atom supercells, with defect-wise mean absolute error below 0.05 eV. In contrast, MLIPs trained on the same dataset can err by more than 1 eV. These results show that charge-density learning enables more robust cross-size transfer than direct energy-force fitting and that mixed-size data design can substantially reduce the cost of defect prediction.}


\maketitle

\section{Introduction}

Defects have a critical influence on the properties of semiconductor materials. Density-functional-theory (DFT) calculations provide a rigorous framework for determining defect formation energies, capture cross sections, transition levels, and related quantities\cite{Freysoldt2014,2016Alkauskas,WEI2004337,weizhang,Van2004}. In DFT calculations, however, isolated defects must be represented in supercells large enough to suppress image interactions\cite{Freysoldt2014,Van2004,Freysoldt2009}. The cost of solving the electronic structure then scales approximately as $O(N^3)$ with system size\cite{Kresse1996}, making high-accuracy defect calculations expensive and limiting the reach of conventional DFT.

In recent years, advances in machine learning have provided a promising alternative. Graph-neural-network (GNN) representations of materials encode local atomic environments and have been used to predict energies, charge densities, and Hamiltonian matrix elements from structure\cite{M3GNet2022,CHGNet2023,Kavanagh2026}. Machine-learning interatomic potentials (MLIPs) can evaluate energies and forces for million-atom systems on a single GPU\cite{Allegro2022,mace2023}. Machine-learning charge densities (MLCDs)\cite{Fabrizio2019,ChargE3Net2024,EAC2026} and machine-learning Hamiltonians (MLHs)\cite{deeph2022,hamgnn2023} further reduce the effective cost of electronic-structure prediction from $O(N^3)$ to nearly $O(N^1)$. These approaches are therefore promising candidates for accelerating defect calculations.

Accurate prediction of defect behavior requires the model to learn the correct local atomic interactions near the defect. However, these interactions converge only in sufficiently large supercells\cite{Freysoldt2014}, so many defect machine-learning models are trained on large-supercell datasets\cite{Kumagai2025,Dutta2026,GaN2022,linderalv2025}.
Furthermore, a supercell usually contains only one defect within a large bulk region, leading to a low defect information density. Therefore, training a model for a single defect typically demands hundreds to thousands of large supercells. For example, Linder\"{a}lv et al.\cite{linderalv2025} used 1341 288--384-atom structures to train an MLIP for the $\text{V}_\text{Si}\text{V}_\text{C}$ defect in 4H-SiC, and Shimizu et al.\cite{GaN2022} used 9100 GaN structures containing 32--128 atoms to train an MLIP for $\text{V}_\text{N}$. Preparing such datasets can take weeks.

Considering the balance among supercell size, cost, and accuracy, we propose that MLCD is a more suitable pathway for this cross-size transfer. First, the charge density is the key variable of DFT: within Kohn-Sham theory, it contains the complete information needed to determine the total energy\cite{Hohenberg,KSeq}. Learning it directly avoids the information loss that occurs when expensive Kohn-Sham calculations are reduced to low-dimensional global targets (i.e., total energy and forces) commonly used by MLIPs. Second, charge density is spatially resolved and interpretable. Local density patterns learned in small cells can be mapped directly to the same local environments in larger cells, provided that the datasets samples the relevant defect and bulk regions. This locality offers a practical criterion for designing more efficient training set.

Here we use MLCD to optimize the tradeoff between formation-energy accuracy and datasets cost. The model is trained for four intrinsic defects in $\mathrm{GaN}$: the nitrogen vacancy $\mathrm{V}_{\mathrm{N}}$, the gallium vacancy $\mathrm{V}_{\mathrm{Ga}}$, gallium on a nitrogen site $\mathrm{Ga}_{\mathrm{N}}$, and nitrogen on a gallium site $\mathrm{N}_{\mathrm{Ga}}$. We partition the defect supercell into 3 spatial regions according to the local charge response and use this analysis to mix defect supercells of different sizes in the datasets. The resulting MLCD model exhibits a progressive cross-size learning behavior distinct from MLIP. With only 96 perturbed configurations spanning 16--96 atoms, it predicts formation energies in 360-atom supercells with mean errors below $0.05$ eV. At fixed accuracy, this mixed-size design reduces the computational cost by a factor of 6.2 relative to training with a single 96-atom supercell size.

\section{Results}

\begin{figure*}[!tbp]
	\centering
	\includegraphics[width=1.0\textwidth]{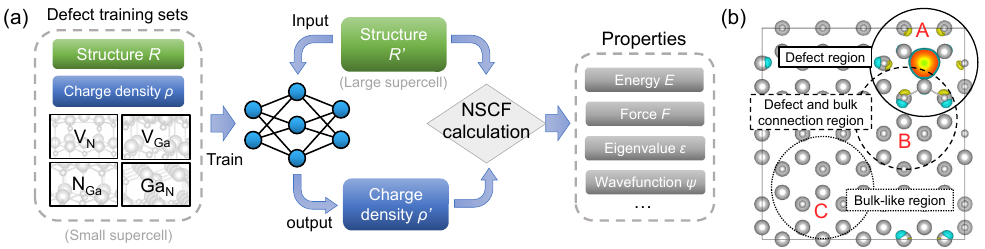}
	\caption{(a) MLCD workflow. Charge-density data from defect supercells are used to train an MLCD model, which predicts charge densities from atomic structures; defect properties are then obtained by NSCF calculations. (b) Spatial partitioning of defect-affected regions inside a supercell.}
	\label{fig1}
\end{figure*}

Figure~\ref{fig1}(a) summarizes the formation-energy prediction workflow. The training set contains DFT structures $R$ and charge densities $\rho$ for four small defect supercells in $\mathrm{GaN}$, together with a limited number of bulk supercells. After training, the MLCD model takes a larger defect-supercell structure $R^\prime$ as input and predicts its charge density $\rho^\prime$. Electronic-structure quantities for $R^\prime$, including the total energy $E$, forces $F$, eigenvalues $\varepsilon$, and wave functions $\psi$, are then obtained from a non-self-consistent-field (NSCF) DFT calculation at fixed $\rho^\prime$. The performance of this MLCD--NSCF force workflow in structure optimization is assessed in Supplementary Information.

\subsection{Training with 96-atom defect supercells}

\begin{figure}[!htbp]
	\centering
	\includegraphics[width=0.6\linewidth]{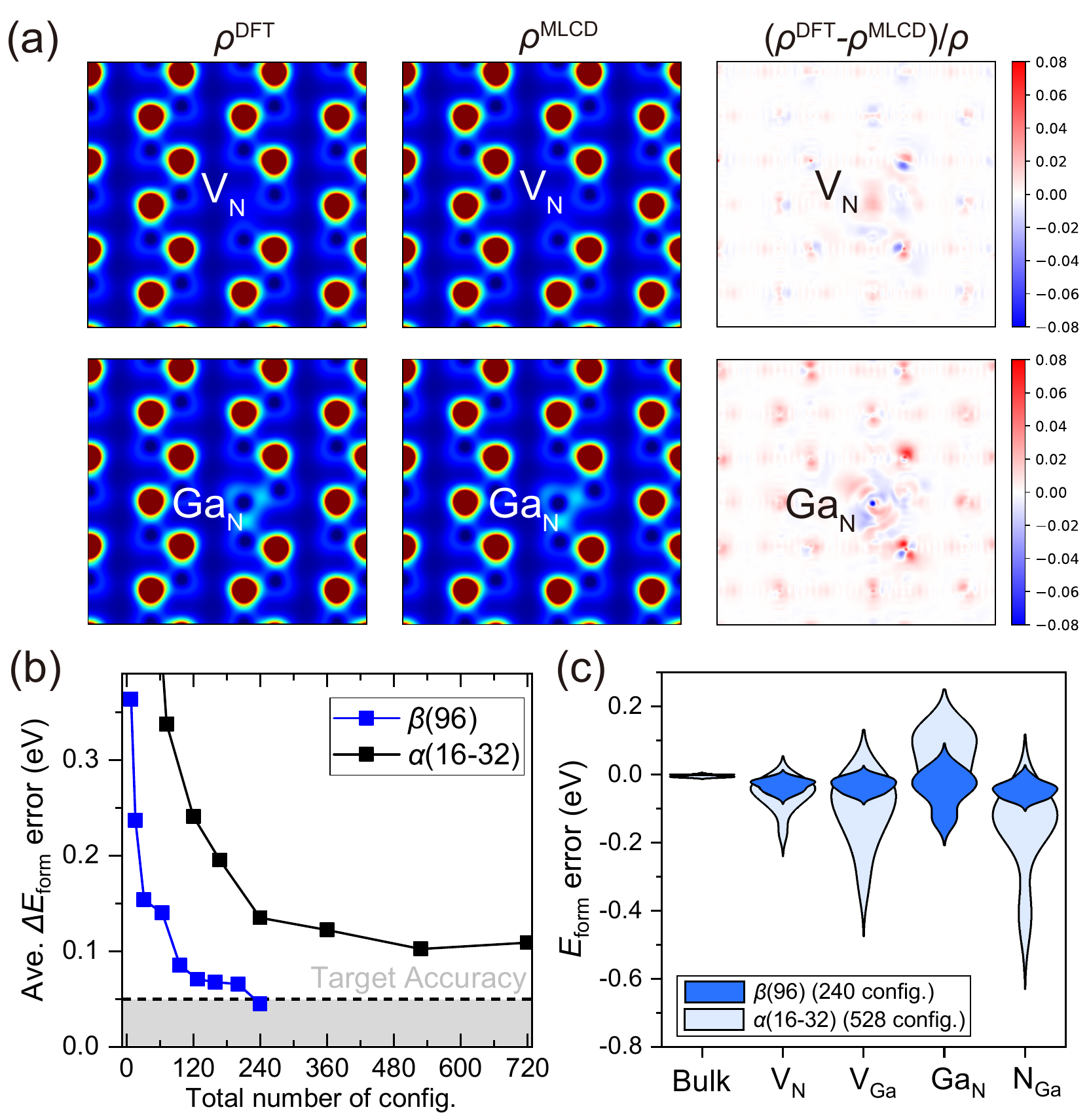}
	\caption{Benchmark results for the 96-atom-supercell training set. (a) Charge-density slices of 360-atom $\mathrm{V_N}$ and $\mathrm{Ga_N}$ defect supercells predicted by the $\beta$(96) model, compared with DFT. (b) Mean absolute formation-energy error $\Delta E_\mathrm{form}$ for four defect types as a function of the number of training structures. (c) Statistical distribution of formation-energy errors for bulk and four defects predicted by $\beta(96)$ and $\alpha(16-32)$.}
	\label{fig2}
\end{figure}

We first evaluated MLCD using a training set built from 96-atom supercells. The $\beta$(96) model contains four intrinsic defects in GaN ($\mathrm{V_N}$, $\mathrm{V_{Ga}}$, $\mathrm{Ga_N}$, and $\mathrm{N_{Ga}}$) and the corresponding perturbed bulk structures, with 360-atom defect supercells used as the common prediction target. Figure~\ref{fig2}(a) compares charge-density slices in the YZ plane for 360-atom $\mathrm{V_N}$ (top) and $\mathrm{Ga_N}$ (bottom) supercells, with the X direction passing through the defect. The three columns show the self-consistent DFT charge density $\rho^\mathrm{DFT}$, the MLCD prediction $\rho^\mathrm{MLCD}$, and the relative difference $(\rho^\mathrm{DFT}-\rho^\mathrm{MLCD})/\rho^\mathrm{DFT}$. We define the charge-density error as
\begin{equation}\label{e_mae}
    \varepsilon_\mathrm{mae}=\frac{\int | \rho^\mathrm{DFT}(\boldsymbol{r})-\rho^\mathrm{MLCD}(\boldsymbol{r}) |d\boldsymbol{r} }{\int | \rho^\mathrm{DFT}(\boldsymbol{r})|d\boldsymbol{r}},
\end{equation}
where $\rho^\mathrm{DFT}$ and $\rho^\mathrm{MLCD}$ denote the DFT and MLCD charge densities, respectively. The errors are $0.013\text{ \%}$ for $\mathrm{V_N}$ and $0.056\text{ \%}$ for $\mathrm{Ga_N}$. The residuals are confined mainly to the nearest-neighbor region of the defect and remain small, demonstrating that MLCD can predict charge densities in large-supercell defect configurations.

Total energies and related quantities were obtained from NSCF calculations using the predicted charge density. The formation energy $E_\mathrm{form}$ of a neutral defect is defined as\cite{Freysoldt2014,WEI2004337}
\begin{equation}\label{formation_energy}
    E_{\mathrm{form}} = E^{\mathrm{defect}}_{\mathrm{tot}} - E^{\mathrm{bulk}}_{\mathrm{tot}} - \sum_i n_i \mu_i.
\end{equation}
Here, $E^{\mathrm{defect}}_{\mathrm{tot}}$ and $E^{\mathrm{bulk}}_{\mathrm{tot}}$ are the total energies of the defect and bulk supercells, $\mu_i$ is the chemical potential of the added or removed element, and $n_i$ is the corresponding change in atom number. In the comparison with DFT, both total-energy terms are obtained from MLCD-based NSCF calculations; the chemical-potential contribution is taken from DFT and cancels in the error $\Delta E_\mathrm{form}$. Figure~\ref{fig2}(b) shows the mean absolute formation-energy error $\Delta E_\mathrm{form}$ of the $\beta$(96) model relative to 360-atom DFT results. The lower axis gives the total number of training structures, and the upper axis gives the number per defect type, following Table~\ref{tab:model_summary}. The gray region marks the target accuracy of 0.05 eV. The error decreases monotonically with training-set size. It exceeds 0.1 eV below 100 structures and reaches the target range only near 240 structures.

Because the random sampling scheme for charge-density grid points introduces training randomness, we performed 12 independent runs with the same training set and hyperparameters. Figure~\ref{fig2}(c) gives the resulting error distributions for the 240-structure $\beta$(96) training set. The bulk errors are centered near zero, showing that the model describes bulk environments accurately. Vacancy defects, $\mathrm{V_N}$ and $\mathrm{V_{Ga}}$, also show relatively narrow distributions with small biases. In contrast, the substitutional defects $\mathrm{Ga_N}$ and $\mathrm{N_{Ga}}$ exhibit broader error distributions, consistent with the charge-density errors in Fig.~\ref{fig2}(a). Even when each defect type has the same number of structures in the training set, the prediction error still depends on the specific defect type because of different degrees of lattice distortion.

The $\beta$(96) results establish the feasibility of MLCD for large-supercell formation energies, but the cost remains substantial: the target accuracy requires self-consistent DFT charge densities for at least 240 96-atom supercells. Smaller supercells (16--32 atoms) also contain one defect and use an analogous perturbative sampling of the potential-energy surface (PES), while being much cheaper to compute. Replacing 96-atom training structures with small-supercell data would therefore reduce the upfront cost. Thus, we next test this limit using only 16--32-atom defect supercells.

\subsection{Training with 16--32-atom defect supercells}

The $\alpha$(16--32) model was trained only on 16--32-atom supercells. Figure~\ref{fig2}(b) shows its mean absolute formation-energy error $\Delta E_\mathrm{form}$ for 360-atom targets as a function of training-set size. At small data volume, $\Delta E_\mathrm{form}$ decreases rapidly, as in the $\beta$(96) model. Above 240 structures, however, the error saturates at 0.1 eV, about twice the target accuracy. Even a 720-structure training set does not reach the target range. Since the $\beta$(96) model reaches this range with 240 structures, the limiting error in $\alpha$(16--32) is not insufficient PES sampling but a systematic PES error caused by the small supercell size.

Figure~\ref{fig2}(c) shows the 528-structure error distribution. Relative to $\beta$(96) [Fig.~\ref{fig2}(c)], the bulk error remains centered near zero, so the small-supercell data still describe bulk charge densities well. The errors for substitutional defects ($\mathrm{Ga_N}$ and $\mathrm{N_{Ga}}$) are larger than those for vacancies ($\mathrm{V_N}$ and $\mathrm{V_{Ga}}$), with individual $\mathrm{N_{Ga}}$ errors reaching 0.6 eV. The defect dependence suggests that the error is relevant to the spatial range of the defect-induced perturbation. Thus, even large amounts of small-supercell data leave non-negligible systematic errors for dilute large-supercell defects because the training structures contain strong defect-image interactions. Inspired by defect-correction methods\cite{Freysoldt2014,Van2004}, even relatively small supercells can yield formation energies approximating those of large cells after correction; in that situation, the small cell provides a rough result, while more accurate correction requires larger cells. We therefore test whether adding medium-size supercells can act as cross-scale anchors while retaining the low cost of the 16--32-atom data.

\subsection{Mixed-size supercell training}

\begin{figure*}[!tbp]
	\centering
	\includegraphics[width=\textwidth]{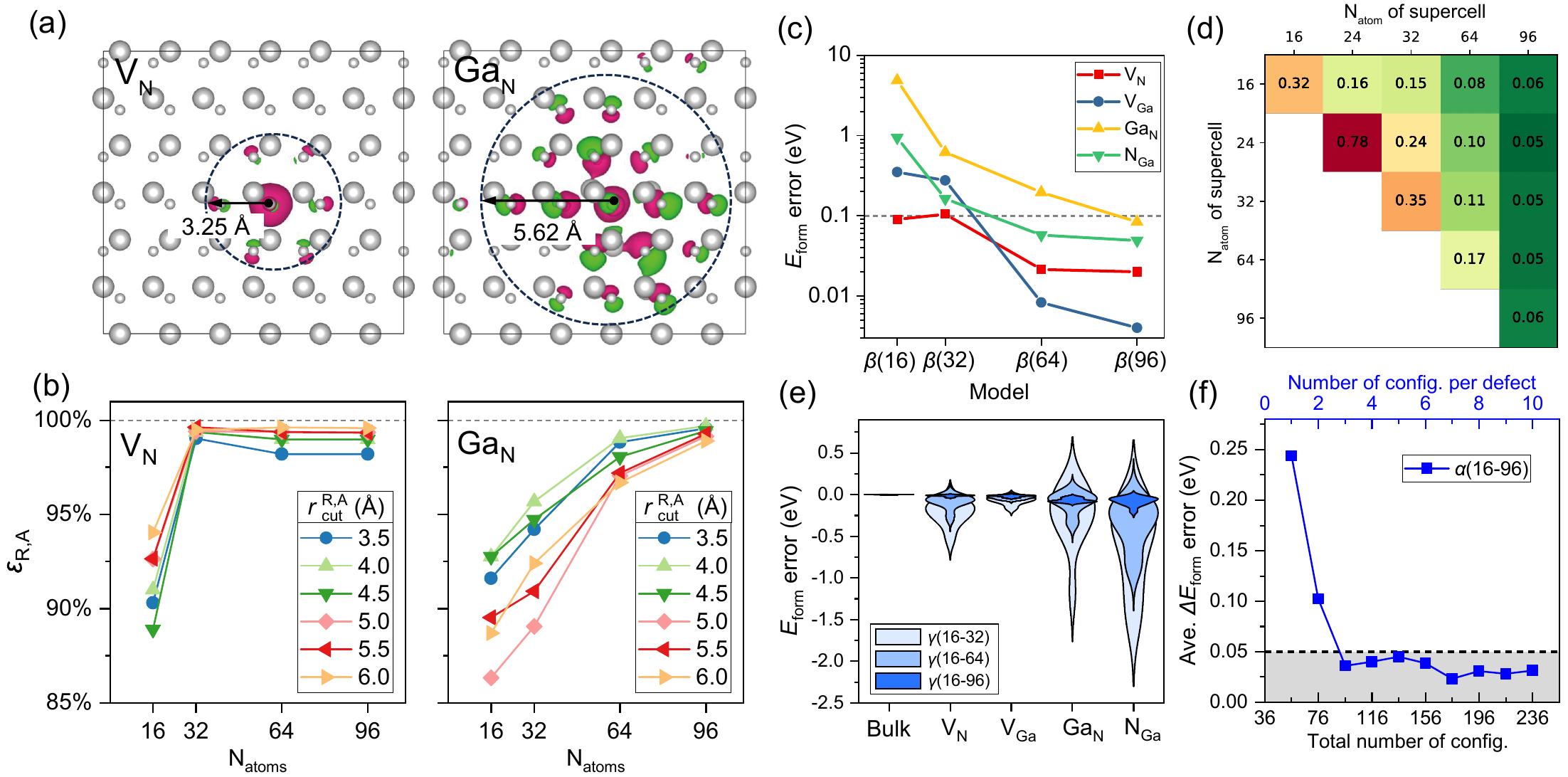}
	\caption{Defect length scale and mixed-size training. (a) Differential charge densities of $\mathrm{V_N}$ and $\mathrm{Ga_N}$ relative to bulk in 360-atom supercells. (b) Structural similarity $\varepsilon_\mathrm{R,A}$ between smaller supercells and the 360-atom reference as a function of cutoff radius. (c) Absolute formation-energy errors for four defects predicted by single-size models $\beta$(16), $\beta$(32), $\beta$(64), and $\beta$(96). (d) Heat map of mean absolute error for different supercell-size combinations. (e) Defect-resolved error distributions for $\gamma$(16--32), $\gamma$(16--64), and $\gamma$(16--96). (f) Convergence of the $\gamma$(16--96) formation-energy error with training-set size.}
	\label{fig3}
\end{figure*}

The defect-dependent degradation in Fig.~\ref{fig2}(c) is closely related to the spatial extent of the defect local charge. Figure~\ref{fig3}(a) shows DFT differential charge densities of $\mathrm{V_N}$ and $\mathrm{Ga_N}$ relative to bulk in 360-atom supercells. For $\mathrm{V_N}$, the charge-density response is confined mainly within a radius of about 3.25 \AA{} and reaches only the next-nearest neighbors, consistent with a localized defect. For $\mathrm{Ga_N}$, the response extends to about 5.62 \AA{} and covers roughly 100 atoms, indicating pronounced non-locality. This contrast explains why the prediction accuracy for $\mathrm{V_N}$ remains acceptable even with training sets of 16 to 32 atoms, whereas $\mathrm{Ga_N}$ is much more sensitive and yields poor predictions.

To quantify how local defect structures vary with supercell size, we introduce the radial distribution function (RDF) $R_i(r)$ and angular distribution function (ADF) $A_i(\theta)$:
\begin{equation}\label{RDF_ADF}
    R_i(r) = \sum_j \delta(r-r_{ij}),\quad A_i(\theta) = \sum_{j,k} \delta(\theta-\theta_{jik}).
\end{equation}
Here, $i$ is the atomic index at the defect position, $j$ and $k$ are the indices of neighboring atoms within the cutoff radius $r_\mathrm{cut}^{R,A}$, and $\theta_{jik}$ is the angle with atom $i$ as the vertex and $\boldsymbol{r}_{ij}$ and $\boldsymbol{r}_{ik}$ as the two sides:
\begin{equation}\label{ADF_theta}
    \theta_{jik} = \arccos\left[\frac{(\boldsymbol{r}_j-\boldsymbol{r}_i)\cdot(\boldsymbol{r}_k-\boldsymbol{r}_i)}{|\boldsymbol{r}_j-\boldsymbol{r}_i|\,|\boldsymbol{r}_k-\boldsymbol{r}_i|}\right].
\end{equation}
On this basis, we define the radial and angular structural similarities as
\begin{equation}\label{error_R_A}
    \begin{aligned}
    S_\mathrm{R}&=\frac{\int R_\mathrm{ref}(r)\cdot R(r)\,dr}
    {\int R_\mathrm{ref}(r)\,dr\int R(r)\,dr},\\
    S_\mathrm{A}&=\frac{\int A_\mathrm{ref}(\theta)\cdot A(\theta)\,d\theta}
    {\int A_\mathrm{ref}(\theta)\,d\theta\int A(\theta)\,d\theta},
    \end{aligned}
\end{equation}
and take their equal-weight average to obtain the total structural similarity,
\begin{equation}
    \varepsilon_\mathrm{R,A}=\frac{1}{2}(S_\mathrm{R}+S_\mathrm{A}).
\end{equation}
We take the 360-atom structure as the reference, where $\varepsilon_\mathrm{R,A}=1$ denotes identical structures. Figure~\ref{fig3}(b) shows $\varepsilon_\mathrm{R,A}$ for smaller supercells at different cutoff radii $r_\mathrm{cut}^{R,A}$. For $\mathrm{V_N}$, the 16-atom supercell ($\bar{a}_{16}/2\approx 2.9$ \AA{}) gives $\varepsilon_\mathrm{R,A}\approx90\%$, indicating that defect-image interactions are still strong. In 32-atom supercells ($\bar{a}_{32}/2\approx 3.42$ \AA{}) and larger, $\varepsilon_\mathrm{R,A}$ exceeds 98\%, indicating that the local $\mathrm{V_N}$ structure is nearly converged. For $\mathrm{Ga_N}$, 16-, 32-, and 64-atom supercells all fail to reproduce the local structure in the 360-atom cell. A 96-atom supercell ($\bar{a}_{96}/2\approx 5.23$ \AA{}) is needed to reach a similarity comparable to that of $\mathrm{V_N}$ in a 32-atom cell. This leads to a key criterion: half of the training-cell lattice length, $\bar{a}/2$, must be comparable to or larger than the defect influence radius. Otherwise, periodic defect-image interactions distort the local structures relative to the true large-cell configurations.

Formation-energy errors confirm this length-scale picture [Fig.~\ref{fig3}(c)]. For the single-size models $\beta$(16), $\beta$(32), $\beta$(64), and $\beta$(96), errors decrease as the training supercell size increases, but the convergence rate is defect dependent. The $\mathrm{V_N}$ error is already below 0.1 eV for $\beta$(32), consistent with its short influence radius. The $\mathrm{V_{Ga}}$ error drops substantially for $\beta$(64), whereas $\mathrm{Ga_N}$ and $\mathrm{N_{Ga}}$ approach 0.1 eV only for $\beta$(96), consistent with the larger spatial response of $\mathrm{Ga_N}$. Thus, accurate large-supercell prediction therefore requires training cells that cover the defect influence radius. Using only large training cells satisfies this requirement but is expensive (Fig.~\ref{fig2}). A more efficient strategy is to mix supercell sizes: a small number of larger cells supply defect-local structure information, while smaller cells provide configurational diversity for bulk-like and transition regions.

We tested this idea by fixing the total number of structures and varying the included supercell sizes [Fig.~\ref{fig3}(d)]. Each grid point $(i,j)$ gives the mean absolute formation-energy error when the training set contains all sizes from $N_\mathrm{atom}=i$ to $N_\mathrm{atom}=j$. Each combination contains about 80 perturbed defect structures and 32 perturbed bulk structures from 16--32-atom supercells, for a total of 112 structures. The diagonal entries are single-size models: the error decreases from 0.32 eV for the 16-atom model to 0.06 eV for the 96-atom model, consistent with supercell-size convergence of defect interactions. Moving right along a row, larger supercells are added while the smallest size is fixed, and the error decreases sharply. Comparing entries in a column shows the complementary role of small cells. For example, with a largest size of 32 atoms, $\{16,24,32\}$ gives 0.15 eV, below both $\{24,32\}$ (0.24 eV) and the single 32-atom model (0.35 eV). Small supercells cannot by themselves support accurate large-supercell prediction, but they supply nonredundant information in mixed training.

The error distributions in Fig.~\ref{fig3}(e) further support this conclusion. As the maximum training size increases from 32 to 96 atoms, all four error distributions narrow, and $\gamma$(16--96) keeps the MAE of every defect within 0.1 eV. Figure~\ref{fig3}(f) shows that $\gamma$(16--96) reaches $\Delta E_\mathrm{form}<0.05$ eV with 96 structures, or about three perturbed structures per defect and per supercell size. By contrast, $\beta$(96) requires about 240 structures to reach the same target. The mixed-size model therefore reduces both the number of configurations and the average DFT cost per configuration.

\subsection{Electronic-structure prediction}

\begin{figure*}[!htbp]
	\centering
	\includegraphics[width=\textwidth]{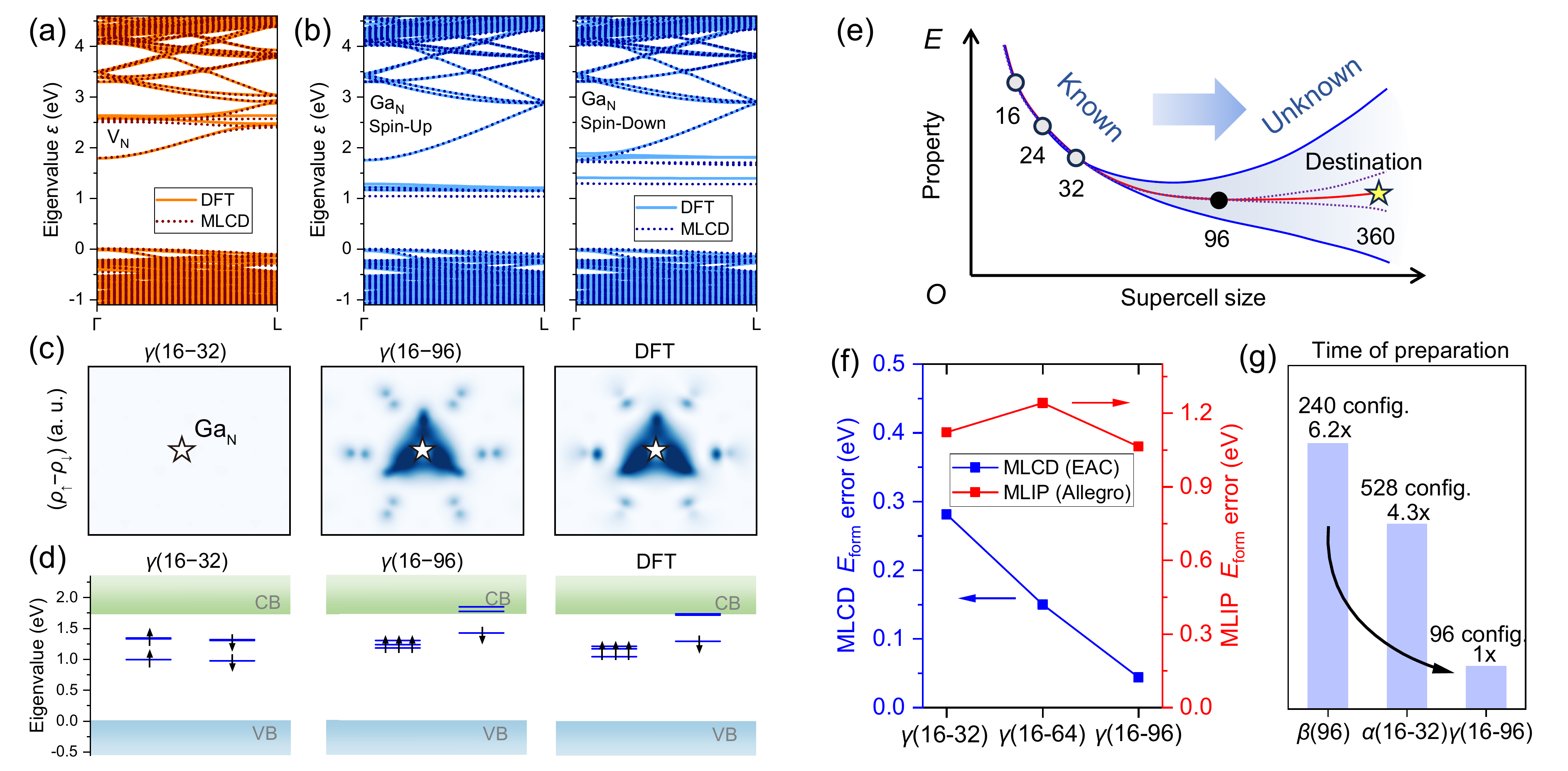}
	\caption{(a) Band structure of a 360-atom $\mathrm{V_N}$ defect predicted by $\gamma$(16--96). (b) Spin-up and spin-down band structures of a 360-atom $\mathrm{Ga_N}$ defect predicted by $\gamma$(16--96). (c) Spin-difference charge density $(\rho_\uparrow-\rho_\downarrow)$ for the 360-atom $\mathrm{Ga_N}$ supercell. (d) Positions and occupations of defect levels in the band gap corresponding to (c). (e) Intermediate-size supercells act as anchors that reduce extrapolation uncertainty from small supercells to the 360-atom target. (f) Effect of the number of supercell-size types on MLCD and MLIP formation-energy errors. (g) Cost under three different dataset strategies.}
	\label{fig4}
\end{figure*}

MLCD-predicted charge densities can also recover electronic-structure properties through NSCF calculations. Figure~\ref{fig4}(a) compares the $\gamma$(16--96) band structure of a 360-atom $\mathrm{V_N}$ supercell along $\Gamma$--$L$ with DFT. Solid and dotted lines denote DFT and MLCD-based results, respectively. The valence-band maximum (VBM) and conduction-band minimum (CBM) are both at $\Gamma$, and the VBM is set to zero. Both DFT and MLCD give a band gap of 1.72 eV. The triply degenerate $\mathrm{V_N}$ defect level at $\Gamma$ lies 2.58 eV above the VBM in DFT and 2.59 eV in MLCD, an error of 0.01 eV. Figure~\ref{fig4}(b) shows the spin-resolved $\mathrm{Ga_N}$ bands. This defect is spin polarized: the lowest spin-up defect level is 1.18 eV above the VBM in DFT and 1.01 eV in MLCD, while the lowest spin-down level is 1.42 eV in DFT and 1.35 eV in MLCD. The single-particle energy error near the VBM is $7\times10^{-5}$ eV for bulk, 0.012 eV on average for $\mathrm{V_N}$ defect levels, and 0.085 eV for $\mathrm{Ga_N}$ defect levels. These results show that MLCD-predicted densities retain sufficient information for accurate large-supercell electronic-structure calculations.

Accurate spin polarization in $\mathrm{Ga_N}$ requires spin-split structures in the training set. Figure~\ref{fig4}(c) compares the spin-difference density $(\rho_\uparrow-\rho_\downarrow)$ in the 360-atom $\mathrm{Ga_N}$ supercell. The $\alpha$(16--32) model predicts nearly zero spin density, whereas $\gamma$(16--96) reproduces the triangular defect-centered spin distribution from DFT. The corresponding gap levels in Fig.~\ref{fig4}(d) show the same trend: $\alpha$(16--32) gives degenerate spin channels, while $\gamma$(16--96) reproduces the DFT spin splitting and level positions. This difference originates from the training data. In supercells with 32 atoms or fewer, strong defect-image interactions suppress spin splitting in relaxed DFT $\mathrm{Ga_N}$ structures, so $\alpha$(16--32) contains no spin-polarization information. In 96-atom supercells, the image interaction is sufficiently weakened and the relaxed structure exhibits the correct spin splitting. Including these configurations allows $\gamma$(16--96) to transfer the spin-polarized defect physics to the 360-atom cell.

\section{Discussion}

\subsection{Mixed-size training as a continuation problem}

The formation-energy and electronic-structure results show that mixed-size supercells improve MLCD transfer to large cells (Figs.~\ref{fig3} and Figs.~\ref{fig4}(a-d)). This behavior can be interpreted as a continuation problem. As sketched in Fig.~\ref{fig4}(e), a property $E$ can be viewed as an unknown function of supercell size $N_\mathrm{atom}$. The function varies rapidly for small cells and approaches a converged value as the supercell grows. Training supercell sizes provide sampled points on this function, whereas the 360-atom target lies outside the sampled region. If only 16--32-atom cells are used, all samples lie in the rapidly varying regime and the continuation path to the target is weakly constrained. Intermediate-size cells, such as 64- and 96-atom cells, provide anchors closer to the converged regime. They shorten the extrapolation distance and constrain the local slope of the size-dependent function, reducing the uncertainty of the continuation. This picture is consistent with the heat map in Fig.~\ref{fig3}(d): diagonal entries, which sample only one size, are poorly constrained, while adding larger sizes along each row progressively lowers the error. The reduction from 0.15 eV for $\{16,24,32\}$ to 0.05 eV for $\{16,\ldots,96\}$ illustrates the effect of a small number of intermediate anchors.

This small-data extrapolation strategy is effective only for MLCD in our tests. We trained an Allegro MLIP on the same $\gamma$ datasets used in Fig.~\ref{fig4}(e) and evaluated 360-atom formation energies [Fig.~\ref{fig4}(f)]. The average MLIP errors remain near 1.0 eV even with the cross-size supercell dataset, far above the MLCD errors. The difference reflects the output representation. MLCD predicts a real-space charge density, so a density pattern learned in a small cell can be transferred to a matching local environment in a large cell. Training data from different sizes therefore impose compatible local constraints. MLIPs instead fit total energies and forces, which are global or force-derived quantities. Data from different supercell sizes do not provide the same local correspondence, limiting cross-scale transfer.

Mixed-size training also weakens the usual monotonic dependence of accuracy on the number of defect supercell datasets. At the 0.05 eV target accuracy, $\gamma$(16--96) uses only 96 structures, compared with 240 structures for the pure 96-atom training set [Fig.~\ref{fig4}(g)]. The structure count is reduced to 40\% of the original value, and the total computational cost decreases by a factor of 6.2. Thus, this mixed-size strategy introduces a novel data augmentation paradigm and offers a highly cost-effective pathway.

\section{Conclusion}

We have developed a mixed-size training strategy for MLCD-based large-supercell defect calculations. For four intrinsic defects in GaN, a model trained only on 96-atom supercells reaches the 0.05 eV target accuracy but requires about 240 DFT training structures. A model trained only on 16--32-atom supercells remains systematically biased: even 720 structures leave the formation-energy error above 0.1 eV. Differential charge-density and structural-similarity analyses show that this limitation is set by the defect influence radius. Half of the training-cell lattice length must be comparable to or larger than this radius; otherwise, periodic image interactions distort the local defect environments relative to the dilute large-supercell limit. Guided by this criterion, the mixed 16--96-atom training set reaches the target accuracy with only 96 structures, reducing the dataset size by about two thirds relative to the single 96-atom strategy. The same MLCD-predicted charge densities also recover band structures and defect levels through NSCF calculations. Interpreted as a continuation problem, mixed-size training supplies intermediate anchors along the supercell-size direction and constrains cross-size extrapolation. This mechanism relies on the real-space charge density, which carries transferable local electronic information. Direct energy-force MLIPs do not provide the same local correspondence and perform poorly on the same small mixed-size datasets. These results show that MLCD combined with mixed-size training offers a data-efficient route for accurate first-principles defect prediction in large supercells.

\section{Methods}

\subsection{Charge-density model}

We used the EAC-Net package\cite{EAC2026}, which implements a deep-learning framework based on an equivariant atom-probe representation. EAC-Net couples atomic and grid information to use charge-density data efficiently. The training data consist of atomic structures $\boldsymbol{R}$ and the corresponding spin-up and spin-down charge densities, $\rho_\uparrow$ and $\rho_\downarrow$. Local environments were constructed for both atomic centers and probe points, with atom--atom and atom--probe cutoff radii set to 6.0 \AA{}. For each training sample, probe points were randomly selected from the same structure. The learning rate decayed exponentially from $3\times 10^{-3}$ to $1\times10^{-5}$.

The EAC architecture preserves rotational equivariance explicitly in a spherical-harmonic basis, with the angular-momentum expansion truncated at $l_{\mathrm{max}} = 3$. Atomic-environment features were extracted with four atomic equivariant convolution layers. The number of basis functions was set to 8, and each layer used 36 feature channels. Atomic invariant features, edge-weight functions, and probe-invariant mappings were implemented with multilayer fully connected neural networks with 128 hidden neurons. The charge-density regression head used two fully connected layers to map the fused equivariant features to a scalar charge density.

\subsection{Training set generation strategy}

Because GNN descriptors act locally within a cutoff radius, transferability to large supercells improves the more closely the target defect configurations match the training data. Based on this assumption, and considering that a large defect supercell contains both defect-like and bulk-like environments, we partition it into three regions [Fig.~\ref{fig1}(b)]. Region A is the defect region, where the defect creates localized electronic and structural changes. Region C is the bulk-like region, far from the defect and only weakly perturbed. Region B connects A and C and covers the transition from defect-perturbed configurations to bulk configurations. Accurate large-supercell prediction requires the training set to sample all three regions.

To sample relaxation-induced distortions in regions A and B, we added perturbed defect and bulk supercells to the training set. All atoms were displaced by random vectors $\Delta R_{j\alpha}$, where $j$ indexes atoms and $\alpha$ indexes Cartesian directions. Within a given supercell, all atoms were assigned the same displacement magnitude to control the perturbation amplitude. The displacement was decomposed as $\Delta \boldsymbol{R}_{j} = R_{\mathrm{disp}} \cdot \hat{\boldsymbol{e}}(\theta,\phi)$, with fixed $R_{\mathrm{disp}}$ and random angular direction $(\theta,\phi)$ for each atom. Thus, each atom was moved from its equilibrium position $R_0$ to a point on a sphere of radius $R_{\mathrm{disp}}$ centered at $R_0$. The supercells in the training set cover multiple perturbation amplitudes, with $R_{\mathrm{disp}}$ linearly distributed from $0\text{ \AA{}}$ to $0.5\text{ \AA{}}$.

Table~\ref{tab:model_summary} summarizes the training set. The Greek letter in each model name denotes a data-preparation strategy. ``Defect supercell $N_\mathrm{atom}$'' gives the supercell sizes included in the training set; for example, $\{16,24,32\}$ denotes $N_\mathrm{atom}=16$, 24, and 32. ``Configs. per $N_{\mathrm{atom}}$ per defect'' gives the number of perturbed structures for each defect and each supercell size. For $\gamma$(16--32), for instance, the training set contains four defect types and three supercell sizes. The $N_\mathrm{atom}=16$ $\mathrm{V}_\mathrm{N}$ subset contains three perturbed structures with $R_{\mathrm{disp}}=0\text{ \AA{}}$, $0.25\text{ \AA{}}$, and $0.50\text{ \AA{}}$. Bulk supercells were perturbed by the same protocol, and ``Total configs. number'' denotes the full training set size.

\subsection{DFT parameters}

DFT data for MLCD training were generated with VASP\cite{KRESSE199615,PhysRevB.59.1758}. The exchange-correlation functional was the Perdew-Burke-Ernzerhof (PBE) \cite{PBE}, and valence electrons were treated with the projector augmented-wave (PAW) method\cite{PhysRevB.50.17953}. The plane-wave cutoff energy was 400 eV. We first optimized the lattice constants of the GaN primitive cell. For bulk supercells, the lattice constants were fixed and self-consistent electronic calculations were performed to obtain charge densities, with an electronic energy convergence threshold of $10^{-5}$ eV. For defect supercells, the lattice constants were allowed to relax, and the force convergence criterion was 0.01 eV/\AA{}. For $N_{\mathrm{atom}}=16$, 24, 32, and 64, the $k$-point density was kept at $a \times N_k \approx 30-40\text{ \AA{}}$, where $a$ is the lattice constant and $N_k$ is the number of $k$ points. We used a $2\times2\times2$ grid for $N_{\mathrm{atom}}=96$ and a single $\Gamma$ point for the 360-atom prediction supercell. All grids were $\Gamma$ centered. NSCF calculations based on MLCD-predicted charge densities used the same computational settings as the DFT training calculations.

\backmatter

\bmhead{Data availability}
The data that support the findings of this study are available from the corresponding authors upon reasonable request.

\bmhead{Code availability}
The EAC-Net package is available at \url{https://github.com/qin2xue3jian4/EAC-Net}.

\bmhead{Supplementary information}
The online version contains supplementary information.

\bmhead{Acknowledgements}
This work was supported by National Key Research and Development Program of China (2022YFA1402904), National Natural Science Foundation of China (12334005, 12188101 and 12404089), Science and Technology Commission of Shanghai Municipality (24JD1400600), and Project of MOE Innovation Platform.

\section*{Declarations}
\begin{itemize}
\item \textbf{Author contribution}: J.Z. performed the calculations, analyzed the data, and drafted the manuscript. M.H. and S.C. conceived and supervised the project. All authors reviewed and approved the manuscript. All authors contributed to writing the manuscript.
\item \textbf{Competing interests}: The authors declare no competing interests.
\end{itemize}

\providecommand{\noopsort}[1]{}\providecommand{\singleletter}[1]{#1}%

\begin{sidewaystableorg}
\centering
\caption{Configuration datasets used for MLCD model training.}
\label{tab:model_summary}
\renewcommand{\arraystretch}{1.15}
\resizebox{\textheight}{!}{%
\begin{tabular}{lccccccc}
\hline\hline
\makecell{Model \\ name}&
\makecell{Defect supercell \\ $N_{\mathrm{atom}}$} &
\makecell{Configs. per \\ $N_{\mathrm{atom}}$ per defect} &
\makecell{Total defect \\ configs. number} &
\makecell{Bulk supercell \\ $N_{\mathrm{atom}}$} &
\makecell{Total bulk \\ configs. number} &
\makecell{Total \\ configs. number} &
\makecell{Corresponding \\ Fig.} \\
\hline
$\beta$(96) & $\{96\}$ & 30 & 120 & $\{96\}$ & 120 & 240 & Fig. 2(c), Fig. 3(c) \\
$\alpha$(16--32) & $\{16,24,32\}$ & 22 & 264 & $\{16,24,32\}$ & 264 & 528 & Fig. 2(c) \\
$\beta$(16) & $\{16\}$ & 30 & 120 & $\{16\}$ & 120 & 240 & Fig. 3(c) \\
$\beta$(32) & $\{32\}$ & 30 & 120 & $\{32\}$ & 120 & 240 & Fig. 3(c) \\
$\beta$(64) & $\{64\}$ & 30 & 120 & $\{64\}$ & 120 & 240 & Fig. 3(c) \\
$\gamma$(16--32) & $\{16,24,32\}$ & 3 & 36 & $\{16,24,32\}$ & 36 & 72 & Fig. 3(e) \\
$\gamma$(16--64) & $\{16,24,32,64\}$ & 3 & 48 & $\{16,24,32\}$ & 36 & 84 & Fig. 3(e) \\
$\gamma$(16--96) & $\{16,24,32,64,96\}$ & 3 & 60 & $\{16,24,32\}$ & 36 & 96 & Fig. 3(e) \\
\hline\hline
\end{tabular}%
}
\end{sidewaystableorg}

\end{document}